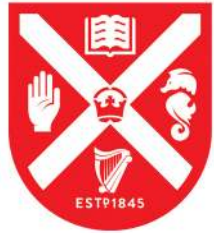

# Excess Electron Localization in Solvated DNA Bases

# Excess Electron Localization in Solvated DNA Bases

Maeve Smyth and Jorge Kohanoff

*Atomistic Simulation Centre, Queen's University Belfast, Belfast BT7 1NN, Northern Ireland, United Kingdom*

We present a first-principles molecular dynamics study of an excess electron in condensed phase models of solvated DNA bases. Calculations on increasingly large microsolvated clusters taken from liquid phase simulations show that adiabatic electron affinities increase systematically upon solvation, as for optimized gas-phase geometries. Dynamical simulations after vertical attachment indicate that the excess electron, which is initially found delocalized, localizes around the nucleobases within a 15 fs time scale. This transition requires small rearrangements in the geometry of the bases.

It is well established that ionizing radiation is capable of inflicting damage to biological matter. This happens through direct absorption and also by indirect effects such as the generation of secondary species and heating of the surroundings [1–4]. One of the most abundant secondary species are low-energy electrons (LEEs) [1–8]. It has been shown that their energy distribution peaks near zero and then decreases monotonically with increasing energy with the majority of the electrons having energies below 30 eV [7–10]. During the past decade there has been increasingly stronger evidence that secondary LEEs ($<20$ eV) have the capability of producing single and double strand breaks in plasmid DNA [6,11–13]. In effect, it is now widely accepted that LEEs play a very important role in radiation damage to DNA. It is thus with this backdrop that recent work has been directed to understanding the processes involved with low-energy electrons and their interactions with biomolecules.

A secondary electron generated in the biological environment by ionization of water, DNA or other biological matter, will travel through the medium losing kinetic energy due to inelastic collisions. The penetration range of LEEs in water is known to be less than 20 nm at energies below 100 eV [14], so that a typical 20 eV secondary electron will travel for about 10 fs until it stops. At this point, or earlier if it hits a resonance, it can attach to DNA. Alternatively, it can create a cavity and localize in water in the form of a *hydrated electron*. A very similar scenario applies to UV ionization and pump-probe experiments [15].

Properties of DNA bases and increasingly large fragments of DNA have been extensively studied in the gas phase as well as in a microsolvated environment. The adiabatic electron affinity (AEA) of a neutral molecule, i.e., the energy difference between the neutral and anionic ground states, provides a measure of the binding energy of an electron. Both theoretical and experimental studies [16–19] agree that the AEA of the isolated pyrimidines (uracil, thymine, and cytosine) are close to zero, while the purines (adenine and guanine) exhibit negative values.

Furthermore, it has been shown experimentally [20] and theoretically [21] that the AEA of the nucleobases increases significantly upon microsolvation. This well-known trend highlights the importance of extending these studies into the condensed phase.

The dynamics of excess electrons in pure water was also the subject of recent computational studies. Boero *et al.* [22] investigated an excess electron in liquid water by first-principles molecular dynamics simulations, finding that after a delocalized period of 1.6 ps, water localizes the electron into a cavity. Similarly, Frigato *et al.* [23] and Marsalek *et al.* [24] examined the surface vs interior localization within a medium-sized water cluster finding that, after a brief delocalized transition period lasting $\sim 2$ ps, the electron localizes around the surface of the cluster. Interestingly, other authors arrived at the seemingly contrasting conclusion that in liquid water the electron is quite delocalized in a 10 Å diameter region, while cavities appear as short-lived fluctuations [25]. The latter are consistent with femtosecond spectroscopy results showing prehydrated electron states with a lifetime of 0.5 ps [15], but contrast with very recent liquid jet photoelectron spectroscopy experiments reporting lifetimes longer than 100 ps and binding energies over 1 eV [26].

In view of the above results, it is natural to wonder what happens to an excess electron in solvated DNA, whether it localizes in the water region or whether it is attracted towards DNA. If the latter scenario was favored, then the question arises of which of the components of DNA will be more attractive for the electron. Understanding the behavior of solvated DNA components due to such electrons is a fundamental step towards modeling DNA radiation damage in a realistic environment. To this end we conducted a computer simulation study of the dynamics of an excess electron in nucleobases immersed in liquid water, as a first step in considering increasingly larger DNA fragments.

We first constructed a model of each system by adding 64 water molecules at random around a central nucleobase [27]. We then carried out a classical molecular dynamics (MD) simulation of the periodically repeated box using the simulation package DL_POLY [28], with the interactions described by the OPLS force field [29]. After equilibrating the system for 1 ns at ambient conditions we ran 1 ns of microcanonical MD and extracted a few representative reference frames. These frames were used as starting points for first-principles MD simulations using the *ab initio* module QUICKSTEP of the CP2K package [30]. The electronic structure was computed within density functional theory (DFT) using the Gaussian and augmented plane waves method (GAPW). In this method the Kohn-Sham orbitals are expanded in a Gaussian basis set, while the Hartree energy and potential are calculated using Fourier transforms. This automatically imposes periodic boundary conditions, thus allowing for condensed phase calculations. For isolated clusters, periodic images were decoupled as in Ref. [31]. Core electrons were replaced by Goedecker-Teter-Hutter (GTH) pseudopotentials [32]. We used fairly complete basis sets, at the triple-zeta plus polarization level (GTH-TZVP), and the charge density was expanded up to an energy cutoff of 250 Ry. The exchange-correlation functional utilized was Perdew-Burke-Ernzerhof (PBE) [33]. It is important to check the effect of electronic self-interaction since the failure of standard functionals to completely remove it may lead to an unphysical spread of the singly occupied molecular orbital. This was done as in [34]. In agreement with previous studies [23], and contrary to the case of positively charged systems, we found that this was not an essential amendment to DFT-PBE calculations for negatively charged ones.

With the aim of extending previous work on microsolvated nucleobases [21], we extracted clusters from the periodic boxes containing an increasing number of water molecules, up to 15, which is significantly larger than the number in previous studies. The first hydration shell contains all the water molecules within 3 Å of each base, which are hydrogen bonded to it. Beyond this radius they are hydrogen bonded to the waters in the first solvation shell. In order to estimate AEAs, we must compare the energies of relaxed neutral and anionic clusters. Instead of optimizing both geometries in the gas phase after extracting the cluster, we have chosen to optimize them in the periodic box, with and without the excess electron. We then extracted the clusters, and did not reoptimize the geometry. We believe that this approach represents the condensed phase more accurately. It has the shortcoming that harmonic zero-point energies (ZPEs) cannot be included as these geometries do not correspond to a stationary point. For fully relaxed clusters ZPE further stabilizes the anion by ~0.1 eV, so that the analysis presented below is not significantly affected. However, it has to be kept in mind that all the AEAs reported here should be shifted upwards in approximately that amount.

We computed the AEA of the clusters at the PBE and the hybrid PBE0 [35] levels using the $6-311++G^{**}$ basis set, which includes diffuse orbitals necessary for anions. PBE and PBE0 results were in agreement, thus confirming that

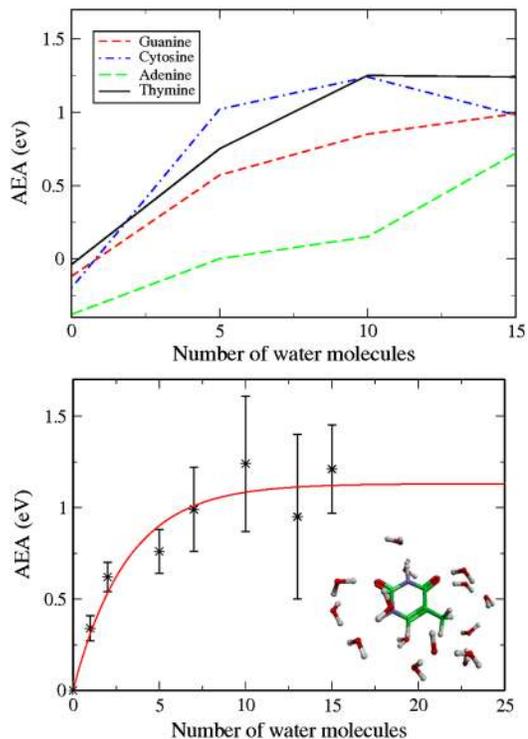

FIG. 1 (color online). Upper panel: AEA of the four microsolvated nucleobases guanine (red, short-dashed line), cytosine (blue, dot-dashed line), adenine (green, long-dashed line) and thymine (black, solid line), as a function of the number of water molecules. Lower panel: averaged adiabatic electron affinity of microsolvated thymine. The solid (red) line is an exponential fit of the calculated data points (stars). The inset shows a typical cluster containing 15 water molecules.

the self-interaction correction is unnecessary in these negatively charged gas-phase systems. In Fig. 1 (upper panel) we present the microsolvation behavior of the AEA for a representative frame of the four solvated nucleobases. These results emphasize the fact that solvation increases the AEA for all the bases.

Figure 1 (lower panel) expands on the above analysis in the case of thymine, by reporting the average of the AEA over 5 representative liquid configurations as a function of the number of water molecules, together with an estimate of their range of variation (in the pictorial form of an error bar). For one and two hydration waters we have computed the affinities in the various possible H-bonding sites finding, in agreement with Kim *et al.* [21], that only binding at the proton acceptor O sites contributes to increasing the AEA. Binding water molecules at the C-H and N-H donor sites has the opposite effect. A sharp increase in AEA is observed upon adding the first two water molecules. The average affinity increases to 0.3 eV and 0.6 eV, respectively, reaching 0.75 eV for the complete first hydration shell and further increasing for 10 and 15 waters. Further solvation has a decreasing influence, thus suggesting that the AEA converges to a value that can be notionally interpreted as the binding energy of an electron to thymine in a bulk water environment. By extrapolating this curve to the bulk limit we obtained a value for thymine of $\approx 1.2$ eV. The ordering of the AEA of the four bases can be rationalized in terms of the number and strength of acceptor sites, which attract electron density when hydrogen bonded. Thymine and uracil are the most favorable ones because they have two (strong) acceptor oxygens. Cytosine and guanine have one O and a weaker N in acceptor positions. In fact, the solvation behavior of their AEA is quite similar, and somewhat less pronounced than in thymine [36]. Adenine has three N acceptor atoms, hence the smaller solvation effect and lower AEA. The emergence of thymine as a slightly more attractive nucleobase in solution, is consistent with recent experimental results on single-stranded oligonucleotide trimers where the most damage was attributed to it [37].

We now turn to the main focus of the Letter, namely, the dynamics of the excess electron. To this end, we simulated the periodic system at constant volume and temperature for an equilibration period of 2.7 ps, and further 0.5 ps at constant energy. At that point we vertically attached an electron and continued the simulation for an additional 0.6 ps. The simulation for the negatively charged system was carried out in the microcanonical ensemble, thus retaining a proper dynamical interpretation while allowing for thermal fluctuations.

The electronic motion was described within the adiabatic approximation, where the electronic density follows instantaneously the nuclear dynamics. In Fig. 2 we show the evolution of the total spin density of the charged system during the initial stages of the MD simulation after electron attachment. We illustrate this for thymine, but the behavior is analogous for all the nucleobases. Initially the excess electron is delocalized over the thymine and water molecules. Within 15–25 fs the electron localizes itself around the thymine as shown in Fig. 2(d), following an adiabatic dynamics driven by the nuclear motion. This is a significant result, showing that under these conditions nucleobases are a very attractive place for the electron to reside. Furthermore, the localization process occurs in such a short time scale that the water cannot reorganize itself and create a cavity to solvate the electron.

To further characterize this process we have examined the total Mulliken charge in the nucleobase orbitals as a function of time after the addition of the excess electron. This is shown in the upper panel of Fig. 3, highlighting that localization is very fast in all cases.

The results of a representative trajectory for thymine are presented in the lower panel of Fig. 3. Initially, less than half of the excess charge is found around the thymine, but within 15 fs around 80% of the charge localizes in the thymine, while the remaining 20% can be accounted for by the neighboring water orbitals. A magnification of the 0–25 fs region is shown in the inset to Fig. 3. By fitting an exponential to this segment of the curve we extracted a time scale for this phenomenon that varied between 5 and 15 fs, depending on initial conditions. After the initial decay the thymine's Mulliken charge fluctuates around

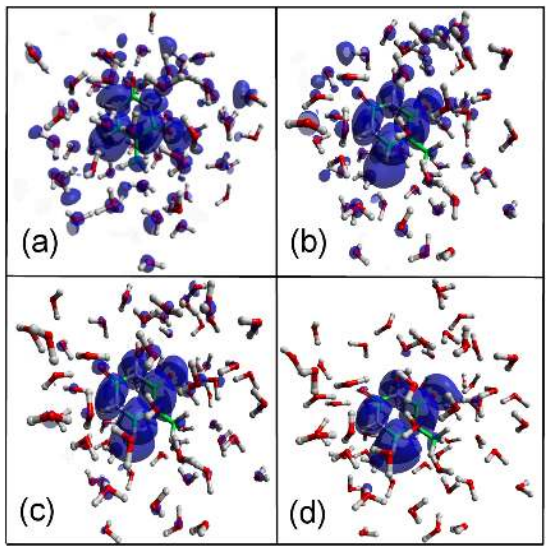

FIG. 2 (color online). Time evolution of the total spin density after vertical electron attachment in solvated thymine, depicted at times: (a) 0, (b) 5 fs, (c) 10 fs and (d) 25 fs. The contour value is always $1 \times 10^{-3} e/\text{Å}^3$.

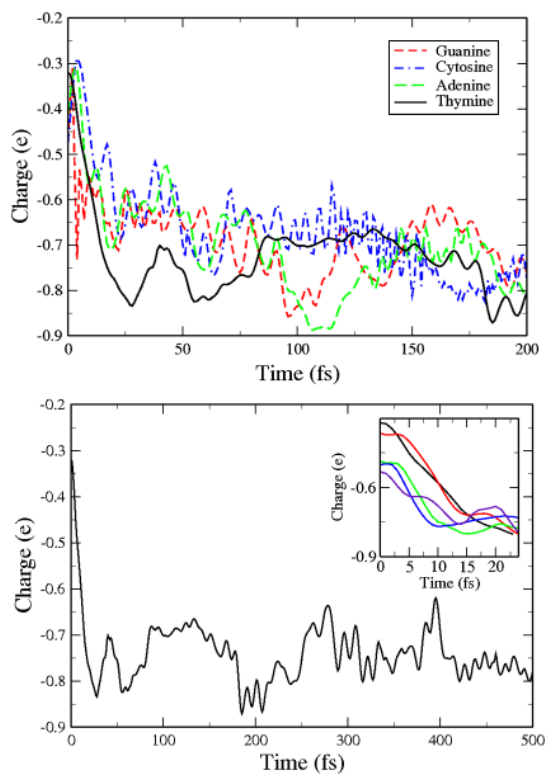

FIG. 3 (color online). Upper panel: Evolution of the Mulliken charge in the four solvated nucleobases (colors and line types as in Fig. 1). Lower panel: Mulliken charge located in the thymine orbitals during one of the MD simulations after vertical attachment of an electron. The inset shows a magnification of the initial behavior, up to 25 fs. The various curves correspond to five independent runs.

$-0.8e$, never to return to the water. It is important to remark that the localization transition, which occurs in all systems studied here, is intimately connected to geometric changes in the nucleobase that are necessary to accommodate the excess electron. The dynamics of these rearrangements dictates the time scale for electron localization.

In building increasingly complex models of solvated DNA, we first considered the sugar group. In the gas phase, individually, neither the sugar nor the base favor electron attachment. Nucleosides, however, do. For example, the AEA of gas-phase thymidine is 0.3 eV. Simulations in the condensed phase show that a vertically attached electron localizes around the nucleobase component in a similar time scale as for solvated thymine (5–15 fs). Therefore, the formation of a glycosidic bond between a sugar C and a base N, has a stabilizing effect on anions that is further enhanced in the aqueous environment.

The above considerations apply to single-stranded DNA. For double strands the first aspect to consider is base pairing. Complementary bases play a similar role to that of water molecules in the first solvation shell, and contribute to increase the AEA from the gas-phase value as shown for the adenine-uracil pair ($\approx 0.4$ eV, up to 0.75 eV in the monohydrate) [38]. In the gas phase the adenine-thymine (AT) pair has a lower AEA than the guanine-cytosine (GC) pair, suggesting that GC is favorable for an excess electron. However, when adding the first hydration shell, the AEAs of AT and CG are very similar ($\approx 0.9$ eV) [39,40]. Our preliminary results for base pairs in liquid water show localization around the pair, but they are not conclusive as of which pair is more attractive, and which base within the pair is favored.

To summarize, we have shown that, in a realistic DNA environment, when a secondary electron has reached low enough energies, it will localize in regions of large electron affinity such as nucleobases in very short times. If this occurs in the vicinity (at least within 2 nm) of DNA, then these electrons will tend to avoid the hydrated state.

We thank T. Youngs for help with Aten and the classical simulations, and F. Currell, D. Timson, T. Todorov, and G. Melaugh, for useful discussions. M. S. thanks N. Forero Martínez for help with quantum chemical calculations. The CP2K calculations were carried out in the HECToR facility under the UKCP consortium allocation.